\documentclass[%
 reprint,
 amsmath,amssymb,
 aps,
prl,citeautoscript,twocolumn
]{revtex4-2}

 \usepackage{amssymb}
\usepackage{graphicx}
\usepackage{amsmath}
\usepackage{bbold}
\usepackage{color}
\usepackage{bm}

\begin{document}
  \title{Stability of Majorana zero modes with quantum optical lattices}

         \author{Santiago F.  Caballero-Benitez}
         \email{ scaballero@fisica.unam.mx}
           \affiliation{Departamento de F\'\i sica Cu\'antica y Fot\'onica, LSCSC-LANMAC, Instituto de F\'\i sica, Universidad Nacional Aut\'onoma de M\'exico, Ciudad de M\'exico 04510, M\'exico}
 \begin{abstract}
 I analyze the emergence of Majorana zero modes (MZM) in a one dimensional ultracold fermionic system confined by an optical lattice inside a high-Q cavity. This forms a quantum optical lattice due to the cavity backaction, with emergent long range interactions controlled by the light pumped into the system and thus long range pairing.  I  investigate the possibility of formation and emergence of edge modes using exact diagonalization and singular value decomposition  of the Hamiltonian in the Majorana representation, while  computing the mass gap of the MZM and its spectral properties. The interplay of cavity induced interactions and  effective $p$-wave short range interactions controls the emergence of the signatures of MZM.  I find that under certain conditions MZM  distinctively emerge leading to a superradiant topological phase of matter. These MZM have potential applications for quantum information as they are topologically protected analogous to the behaviour of the Kitaev chain. The techniques employed lead to a scheme based on the Majorana representation and exact diagonalization with self consistent conditions that can be applied to other fermionic  and analog systems termed Light-Matter-Majorana. 
 \end{abstract}              
    \maketitle
    
    \begin{figure}[t]
    \centering
    \includegraphics[width=0.48\textwidth]{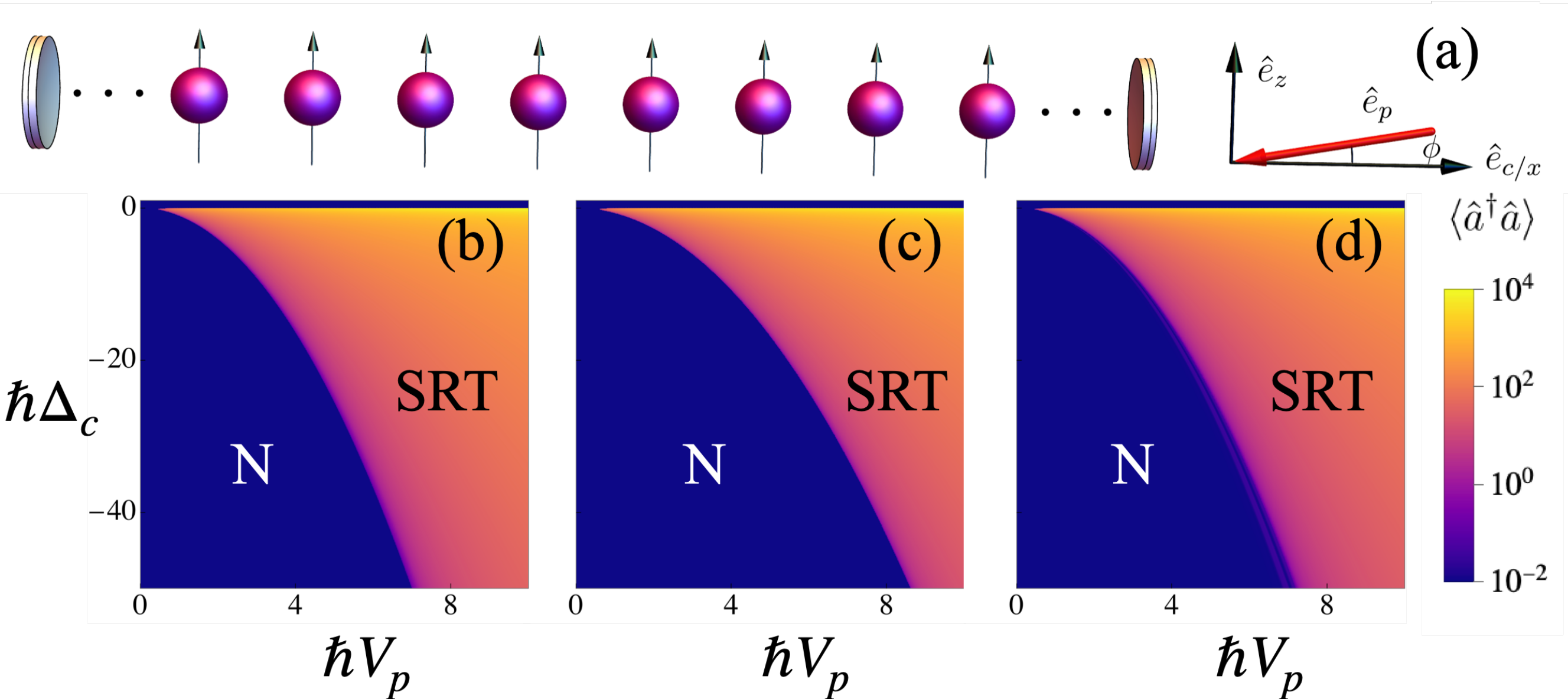}
    \caption{{\bf System setup scheme and phase diagram.} (a) Spin polarized fermions along an optical lattice inside a high-Q cavity.  Light is pumped into the system from the side with in the direction $\hat{e}_p$, the optical lattice is along the $\hat{e}_x$ and the angle of the between them is small $\phi\lesssim 5^{\circ}$. (b-d) Phase diagram with the normal phase (N) and superradiant topologial phase (STR), the bar shows the number of photons in the cavity. The system in the STR besides from collective photon emission supports edge modes. Parameters are: (b) $g_p=0$, $\tilde{\mu}/2\tilde{t}=0$; (c) $g_p=0$, $\tilde{\mu}/2\tilde{t}=0.5$;  (d) $g_p=0.3\tilde{t}$,  $\tilde{\mu}/2\tilde{t}=0$; with $\kappa/\Delta_c=0.3$, $\hbar V_p$ and $\hbar\Delta_c$ are in units of $\tilde{t}$}
    \label{fig:system}
\end{figure}

Recent experimental developments with ultracold fermions inside high-Q cavities~\cite{CF1,CF2,CF3,CF4,CF5}, motivate further analysis of the properties one might be able to control in these systems. Tailoring different regimes and phases of correlated quantum matter with bosonic realizations have been explored with interesting competition scenarios and the emergence of exotic phases of matter with the addition of optical lattices~\cite{E1,E2,E3,H1,P}. Ultracold Fermi systems inside cavities have been explored theoretically with ~\cite{PRAF, FQOL1,FQOL2,FQOL3} and without optical lattices~\cite{WOL1,WOL2,WOL3}. Further, in the advent of investigating fundamental properties of fermionic systems and possible applications for quantum information of Majorana edge modes~\cite{QIA1,QIA2,QIA3} or analog quantum simulation~\cite{AQS} of condensed matter light-matter hybrid systems motivates this work. As it is well known, analog superconducting systems in the tight banding limit with short range pairing present ($p$-wave symmetry)  can support topologically protected Majorana zero modes (MZM) at the edges of the system, the Kitaev model~\cite{Kitaev}.  These MZM are robust towards decoherence as they are topologically protected, and  possibly robust to engineer q-bits with them.

With the above context in mind, I explore the fate of MZM in a fermionic system inside a high-Q cavity where the cavity back action induces effectively long range pairing while the effect of short range interactions is also considered. Thus, the light pumped into the system in combination with the cavity induced dynamics and the lattice can be used to induce emergent MZM triggered by the light-matter coupling. However, the  long range character of the interaction could suppress the emergence of the modes. I explore the possibility to modify the suppression effect in the MZM in the cavity system.  In the past, other finite range interacting models have been studied in connection with the emergence of MZM~\cite{Pupillo, Pupillo2,Lew} portraying a complicated landscape.  I find that indeed MZM can arise and with the interplay of short range interactions these can be stabilized and their amplitude maximized.

    \begin{figure}[t]
    \centering
    \includegraphics[width=0.48\textwidth]{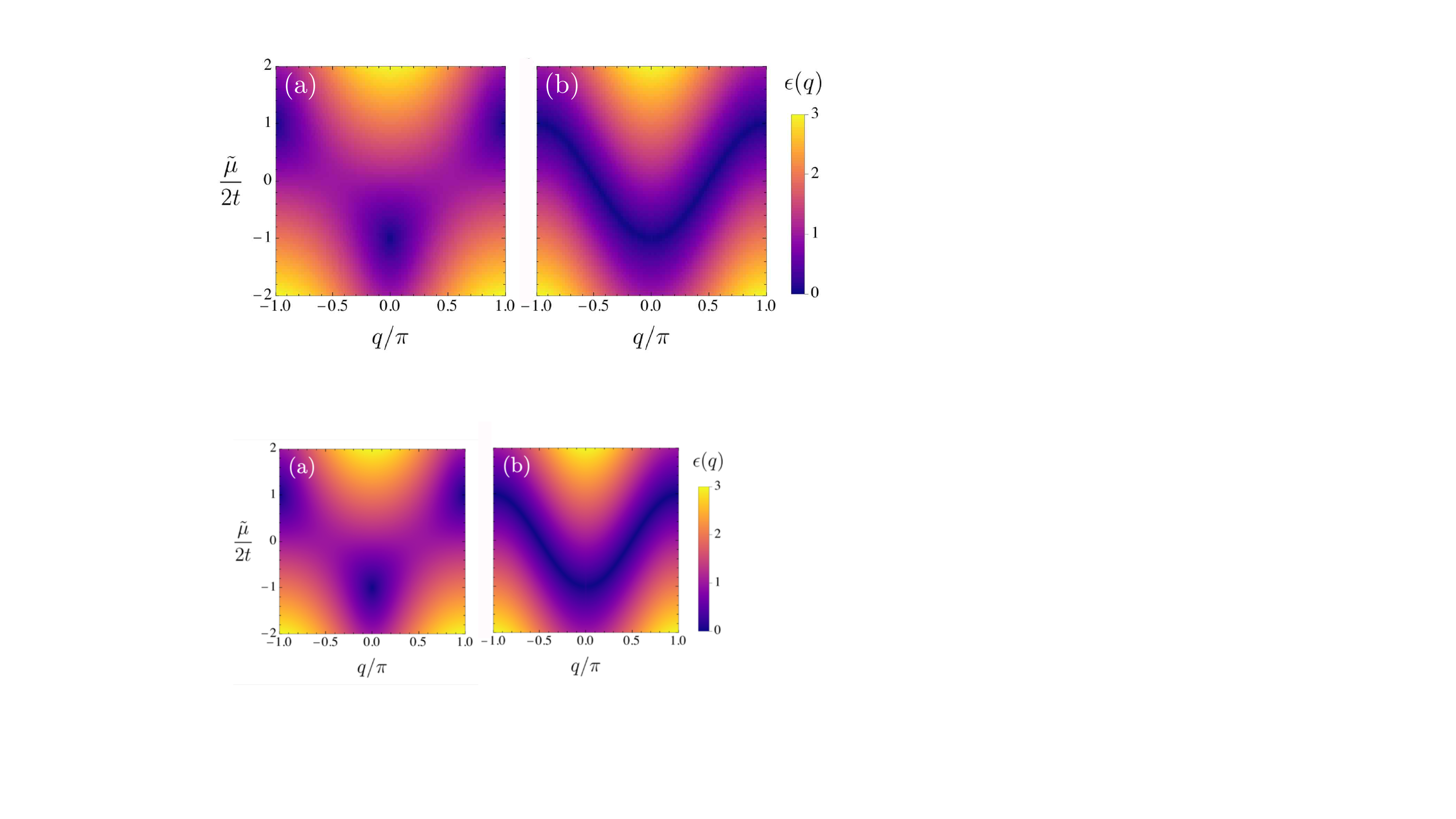}
        \caption{{\bf Bulk dispersion relations}. (a) the Kitaev chain (without cavity, $g_p\neq 0$, $V_p=0$).  (b) the system with only cavity induced interactions ($g_p=0$, $V_p\neq 0$). Parameters are: $N_s=10^3$  with $\epsilon(k)$ in units of $\tilde t$.}
    \label{fig:comp_dis}
\end{figure}

{\it The model of ultracold fermions in a high-Q cavity}. I consider the effective light-matter model of a one dimensional elongated chain of ultracold spin polarized fermions inside a single mode high-Q cavity. Following the methods in~\cite{P,P1}, the low energy light-matter Hamiltonian is
\begin{eqnarray}
\mathcal{H}&=&-\hbar\Delta_c\hat a^\dagger\hat a-t\sum_i(\hat c_i^\dagger\hat c_{i+1}^{\phantom{\dagger}}+\mathrm{H.c.})-g_p\sum_{i}\hat n_i\hat n_{i+1}
\nonumber\\
&+&\frac{\hbar V_p}{\sqrt{N_s}}\sum_i\tilde{g}_i\hat n_i(\hat a^\dagger+\hat a)-\mu\sum_i\hat n_i
\end{eqnarray}
with the number of sites $N_s$. The first term is the cavity energy with  the cavity-pump detunning $\Delta_c=\omega_c-\omega_p\sim\mathrm{GHz}$. The second term is the kinetic energy with tunneling amplitude $t$, the third term is a short-range nearest neighbor interaction with strength $g_p\geq0$  and the last term is the light-matter interaction between the atoms ($\hat c$) and the cavity mode ($\hat a$) with pump strength $V_p$. I assume the fermions (spin polarized, effectively spinless) are trapped by an optical lattice inside a high-Q cavity. The short range {\it p}-wave type interactions  has been included, as on site interactions are Pauli blocked. This nearest neighbor interaction can arise effectively via {\it p}-wave resonances\cite{PWAVE0, PWAVE1, PWAVE} or mediated by bosonic fields\cite{pwave1, pwave2} or a similar effect could be engineered in groups of sites by means of using additional light induced modes, thus changing the geometric arrangement of the light pumped into the system, the cavity axis and the optical lattice~\cite{P1,PRAG}.  The optical lattice potential depth can be modified independently of the cavity pump modulating tunnelling $t\sim\hbar\mathrm{kHz}$ of the atoms and these are in  the lowest Bloch band. I consider that the lattice is deep enough that only density coupling between light and matter is relevant~\cite{P1}. The amplitudes $\tilde{g}_i$  are the mode overlaps between the cavity mode and the spatial variation of the fermionic density, $V_p$ is in units of the Rabi frequency that emerge due to atom-photon transitions (proportional to the pump intensity). In this treatment,  the transitions to excited atomic states have been adiabatically  eliminated  and the Wannier functions of the lowest band corresponding to the optical lattice potential have been implicitly introduced\cite{P}.  In Fig. \ref{fig:system} the setup of the system is shown, where I consider the light is pumped from the side in the same plane as the lattice, generating homogenous coupling at small angle $\phi\lesssim 5^\circ$ see Fig. \ref{fig:system} (a), the smaller the angle the better to approximate homogenous coupling. However, other pump angle choices will lead to additional physics with broken symmetry phases, i.e. density waves~\cite{P,CF1,CF2,CF3,CF4,CF5} or other exotic phases\cite{P1,PRAG,GDI}, which can be considered in principle. In a realistic ultracold matter experiment with a lattice typically the order of sites ranges from $10$ to $10^4$ sites depending on the effective dimensionality of the system and the system always has open boundary conditions, as  I consider here.

{\it Majorana representation}. I integrate out the light degrees of freedom via Hilbert space rotations~\cite{NJP2015}, leading to the effective matter Hamiltonian with long range interactions induced due to cavity back action. It follows to use  the Majorana fermion mapping~\cite{Kitaev} with Majorana fermion operators given by $a_{2j-1}=e^{-i\theta/2}c^{\phantom{\dagger}}_j+e^{i\theta/2}c^{\dagger}_j$, $a_{2j}=(e^{-i\theta/2}c^{\phantom{\dagger}}_j-e^{i\theta/2}c^{\dagger}_j)/i$, where $\theta=\arg(\tilde{\Delta}_{p/l})$ is the phase of the order parameter. In Majorana language,

\begin{eqnarray}
&\mathcal{H}_{\mathrm{eff}}&=-\frac{i\tilde{\mu}}{2}\sum_{j=1}^{N_s} a_{2j-1}a_{2j}
\\
&-&\frac{i\tilde{t}}{2}\sum_{j=1}^{N_s-1}( a_{2j}a_{2j+1}- a_{2j-1}a_{2j+2})\nonumber
\\
&+&\frac{i|\tilde\Delta_p|}{2}\sum_{j=1}^{N_s-1}( a_{2j}a_{2j+1}+a_{2j-1}a_{2j+2})\nonumber
\\
&+&\frac{i |\tilde\Delta_l|}{2 N_s}\sum_{j=1}^{N_s-2}\sum_{k=0}^{N_s-j-1}(a_{2j}a_{2(j+k)+1}+a_{2j-1}a_{2(j+k)+2}).
\nonumber
\end{eqnarray}
Above it has been assumed the geometric condition of the light pumped into the system that generates constant light-matter overlaps ~\cite{PRAW}, $\tilde{g}_i=\tilde{g}$ (Diffraction maxima configuration) and later I have performed the standard mean-field decoupling (Hartree-Fock-Bogoliubov theory) of the fermionic atoms.  The cavity back-action effectively induces long range paring via the long range emergent density-density interaction mediated by the photons, similar to ~\cite{PRAF}.  The effective long range gap is  $\tilde{\Delta}_l=\hbar\tilde{g}_{\mathrm{eff}}\Delta_l$, the effective $p$-wave gap is $\tilde\Delta_p=g_p\Delta_p$, and the effective chemical potential $\tilde{\mu}=\mu-\hbar\tilde{g}_{\mathrm{eff}}\langle\hat n\rangle/2$. The gaps are approximated with their mean value across the system as $\Delta_l\approx\sum_{i<j}\langle \hat c_i\hat c_{j}\rangle/N_s$, $\Delta_p\approx\sum_i\langle \hat c_i\hat c_{i+1}\rangle/N_s$,  and the filling factor is $\langle\hat n\rangle$. The effective chemical potential and the effective long range gap depend on the light pumped into the system via   $\tilde g_{\mathrm{eff}}=\Delta_cV_p^2\tilde{g}^2/(\Delta_c^2+\kappa^2)(1+\kappa^2/\Delta_c^2)$, with the cavity decay rate given by $\kappa$ which is phenomenologically introduced and the term $\kappa^2/\Delta_c^2$ is the non-adiabatic correction\cite{NJP2015,SQOL}. I assume the limit where $|\Delta_c|>\kappa$~\cite{P}. The Fock shifts due to the short range and cavity interactions have been implicitly incorporated in the renormalized effective tunneling amplitude $\tilde{t}$ and the $\tilde{\Delta}_{p/l}$ definitions, full details will be discussed elsewhere. Numerical experiments are performed to study the system  solving the self consistent equations of the mean field treatment  using a parallel GPU implementation, this allows to explore efficiently the parameter space. The treatment is similar to the recently developed Light-Matter-DMRG framework\cite{LMDMRG,GDI}, but here the method has been tailored to treat simultaneously the Hartree-Fock-Bogoliubov self-consistency with exact diagonalization in the Majorana representation of the underlaying matter Hamiltonian and the light amplitude in the full quantum representation, I call this scheme Light-Matter-Majorana. In addition, open boundary conditions with a finite number of sites are used to resemble a realistic experimental situation with ultracold atoms and to take into account the effect of the MZM in the computations. The typical phase diagram is shown in Fig.\ref{fig:system} (b) for $g_p=0$ ($\tilde{\mu}/2\tilde{t}=0$),  (c) $g_p=0$ ($\tilde{\mu}/2\tilde{t}=0.5$), and (d) $g_p=0.1$  ($\tilde{\mu}/2\tilde{t}=0$). The system exhibits normal non-topological phases $\mathrm{N}$ and superradiant topological phases $\mathrm{SRT}$, as I will explain in what follows.

 \begin{figure}[t!]
    \centering
    \includegraphics[width=0.42\textwidth]{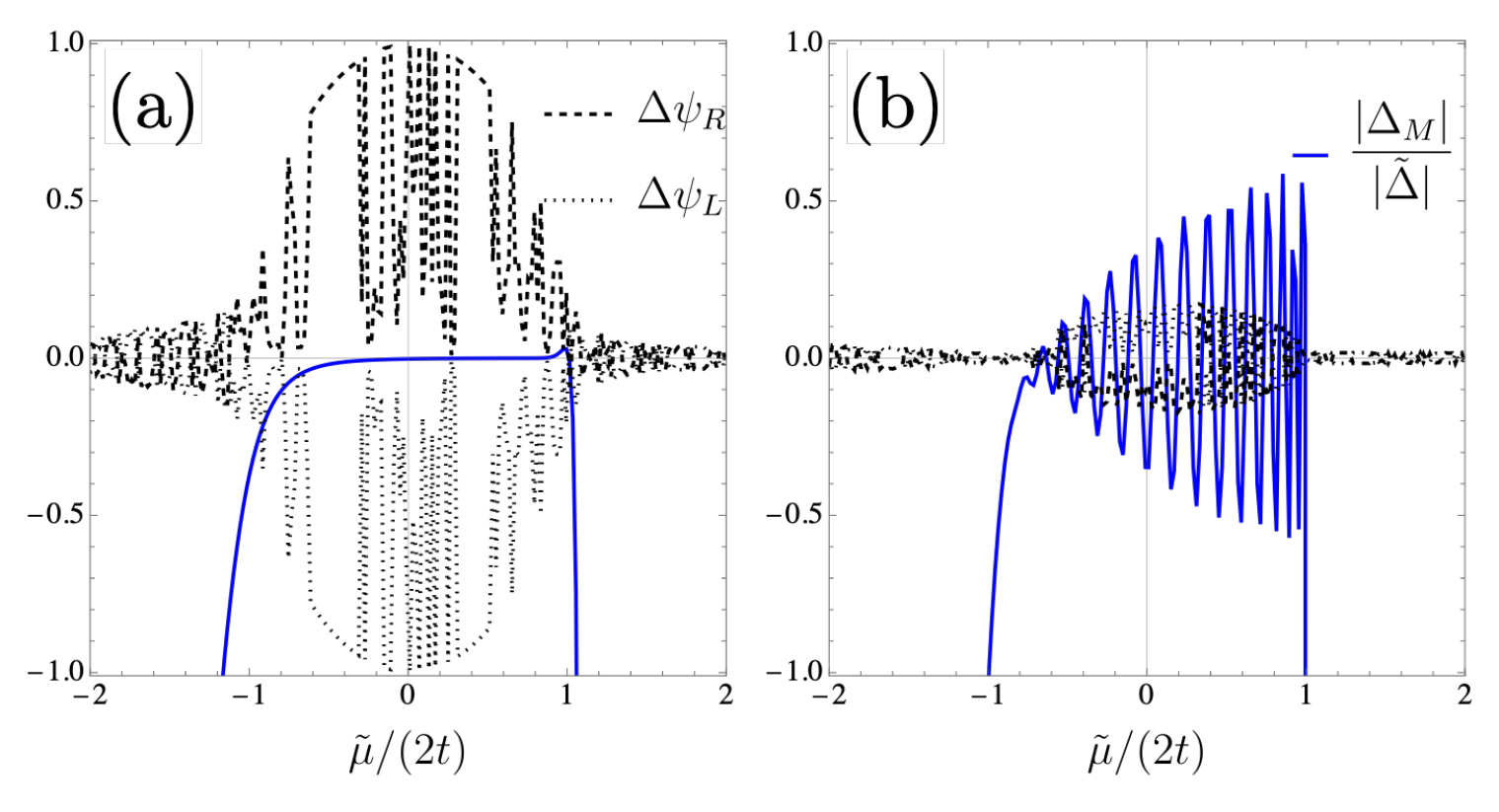}
    \caption{Mass indicator $\Delta_M$ and differences in amplitudes of Majorana modes at the left/right edges $\Delta\psi_{L/R}$.  (a) $V_p\neq0$ $g_p\neq0$. (b)$V_p\neq0$ $g_p=0$ For (a), MZM are present with large amplitude for $\tilde{\mu}/2\tilde{t}<1$, $|\Delta\psi_{L/R}|\sim 1$. For (b), MZM have suppressed amplitudes.  In (b) the mass indicator fails to predict the location of the MZM, as it oscillates. Parameters are:  $N_s=40$ and $|\tilde{\Delta}_l|=2\tilde{t}$.
    }
    \label{fig:dpsi_cavs}
\end{figure} 

{\it  Considering only cavity induced interactions $g_p=0$}. The bulk dispersion of the system is given by $\epsilon(q)=\pm\sqrt{(\tilde\mu+2\tilde{t}\cos(q))^2+4|\tilde\Delta_l|^2f(q)^2}$  for $-\pi\leq q\leq \pi$, with $f(q)=\sin(N_s q/2)\sin((N_s+1)q/2)/(N_s\sin(q/2))$~\cite{limit}.
The dispersion relation for the Kitaev chain and $\mathcal{H}_{\mathrm{eff}}$ are shown in Fig. \ref{fig:comp_dis}. The Kitaev chain has only three Dirac points for $|\tilde{\mu}/2\tilde{t}|\leq1$ at $q=0$ ($\tilde{\mu}/2\tilde{t}=-1$) and $q=\pm\pi$ ($\tilde{\mu}/2\tilde{t}=1$).  In contrast, for $\mathcal{H}_{\mathrm{eff}}$  there is a continuum of Dirac points starting a $q=0$ ($\tilde{\mu}/2\tilde{t}=-1$) and opening in pairs for different values of $q$ [$q\approx\arccos(-\tilde{\mu}/2\tilde{t})$] up to the $q=\pm\pi$ pair ($\tilde{\mu}/2\tilde{t}=1$) for large $N_s$. This occurs because the cavity induced effective gap amplitude $|\tilde{\Delta}_lf(q)|$  has its maximum values near $q=0$ and otherwise it almost vanishes with oscillatory behaviour.  Being more precise, for a finite system there are $2N_s+1$ Dirac points in the  interval  $|\tilde{\mu}/2\tilde{t}|\leq1$. The Dirac points are,
\begin{equation}
q^D_{\mathrm{cav}}(n)=\pm\frac{(-1)^n(N_s-n+1)+(2 n-1) N_s+n-1}{2 N_s (N_s+1)}\pi
\end{equation}
 with $n\in\mathbb{Z}^+$ and  $n\leq N_s+1$. Starting from $q=0$ the Dirac points open in symmetric pairs as shown in  Fig. \ref{fig:comp_dis} (b). For finite $N_s$, when $\tilde\mu=-2\tilde{t}$, at the Dirac point with $q=0$ the dispersion is massless, $\epsilon(q)\approx \pm v_F(N_s+1) |q|$, with $v_F=|\tilde\Delta_l|$. The origin of this Dirac point is precisely the particular form of the double peak structure of $|f(q)|^2$.  In the opposite limit  $\tilde\mu=2\tilde{t}$, the dispersion is also massless, with $\epsilon(q\mp \pi)\approx \pm v_F|q|$.
In other finite range models ~\cite{Pupillo},  one recovers an equivalent dispersion for the Dirac points at $q=\pm\pi$, but the one at  $q=0$ is absent.
In general near the $2N_s+1$ Dirac points  in $\mathcal{H}_{\mathrm{eff}}$, I have that the dispersion is
\begin{figure}[t!]
    \includegraphics[width=0.42\textwidth]{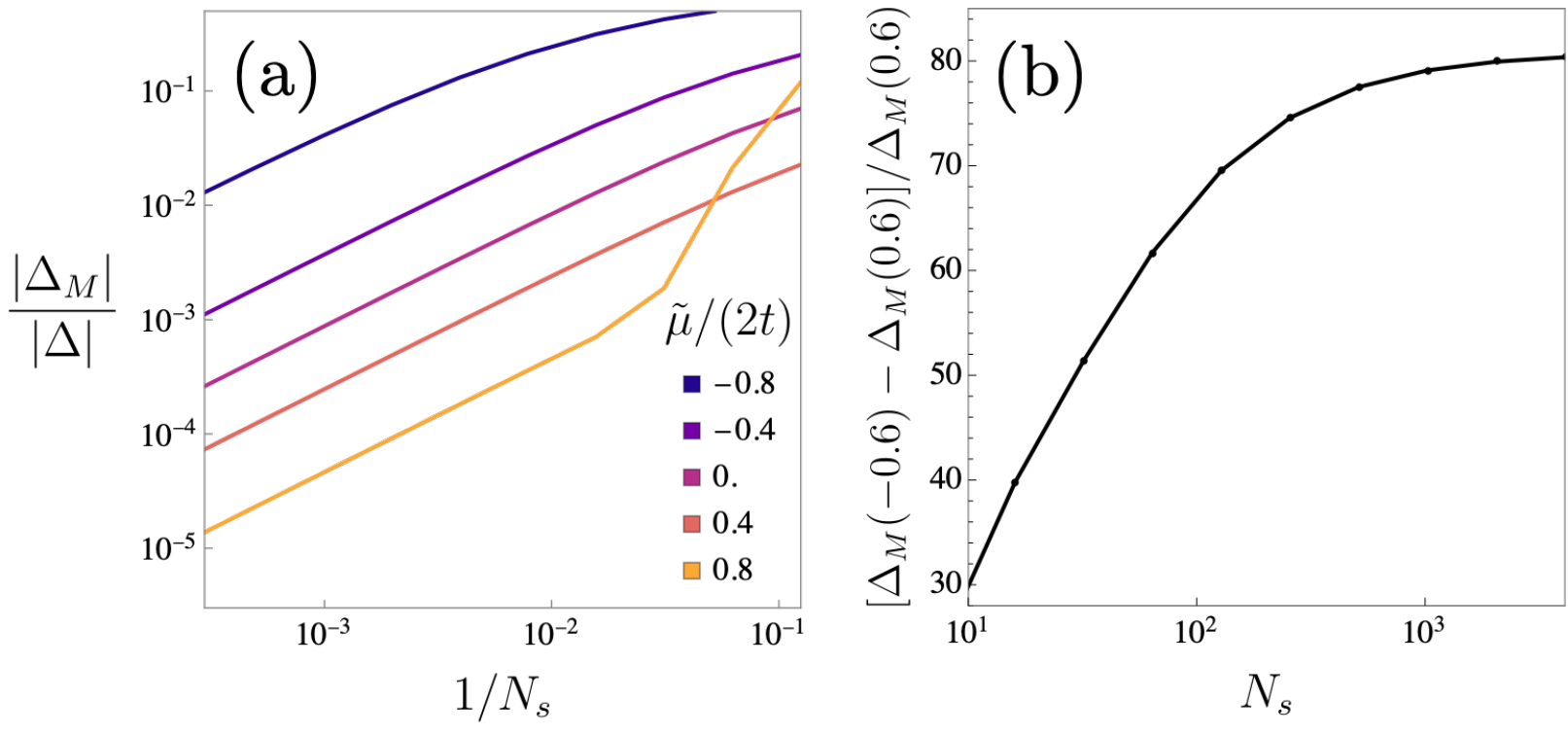}
      \caption{(a) Mass indicator $\Delta_M$ of the MZM, with $|\tilde{\Delta}_l|/2\tilde{t}=1$ for $g_p\neq0$. The MZM are localized at the edges as the number of sites $N_s$ increases. (b) Speed factor difference between the convergence of  the MZM $\Delta_M$ at $\tilde{\mu}/2\tilde{t}=-0.6$ and $\tilde{\mu}/2\tilde{t}=0.6$.
    }
    \label{fig:massind}
\end{figure}

\begin{equation}
\epsilon^D_n(q)\approx \pm \frac{(N_s+1)v_F |q|}{N_s+\frac{1-(-1)^n}{2}}\stackrel{N_s\gg 1}{=}\pm v_F |q|
\end{equation}
for each $q^D_{\mathrm{cav}}(n)$ for $2\leq n\leq N_s+1$ and for $q^D_{\mathrm{cav}}(1)=0$, $\epsilon^D_1(q)\approx \pm v_F(N_s+1) |q|$.

The Dirac points at   $|\tilde{\mu}/2\tilde{t}|=1$ in $\mathcal{H}_{\mathrm{eff}}$ suggest that the system with the cavity induced interaction supports zero modes with paired Majorana fermions at different sites as long as $2|\tilde{t}|>|\tilde{\mu}|$.
 In what follows, I investigate if this is the case fixing the gap parameters $\tilde\Delta_{p/l}$ without loss of generality. It is worth noting that only for finite $N_s$ and open boundary conditions we will expect to have MZM. Thus, ultracold experiments are ideal for this task, as system size can be tailored in principle.
 
{\it The fate of Majorana edge modes.} The Hamiltonian without light matter interactions ($V_p=0$) and $g_p\neq 0$ supports Majorana edge modes as shown by Kitaev\cite{Kitaev}. Thus, I explore the situation $V_p\neq 0$ and $g_p\neq0$. The emergence of the MZM's can be studied with methods similar to ~\cite{Pupillo2}, following the technique of ~\cite{Fendley}. I introduce the mass gap ($\Delta_M$) parameter that quantifies effectively the amplitudes in the single particle states at different sites from the edges at the level of the commutation algebra with the Hamiltonian.

\begin{figure*}
    \centering
    \includegraphics[width=0.98\textwidth]{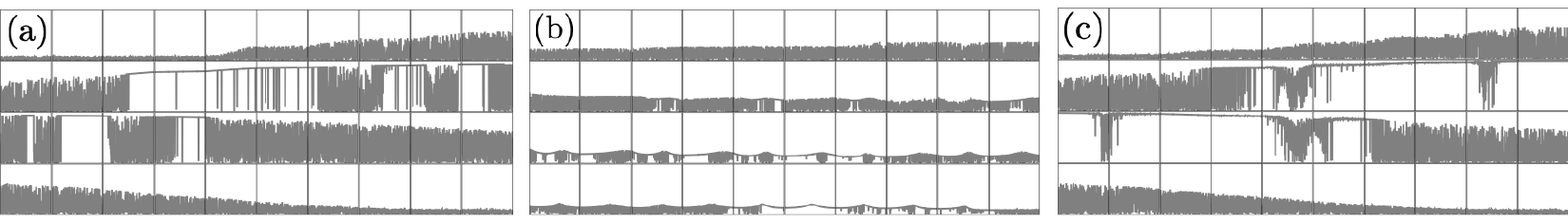}
    \caption{
  {\bf Majorana windows}. The effective amplitude of edge modes $D\psi= ||\Delta\psi_{L/R}|-\bar\psi_{L/R}|$  for (a) $V_p\neq0$ $g_p\neq0$, (b)  $V_p\neq0$ $g_p=0$ and (c) $V_p=0$ $g_p\neq0$ the Kitaev chain. Each panel corresponds to an interval $[\tilde{\mu}^D_n,\tilde{\mu}^D_{n+1}]$ with the height of each panel in the interval $[0,1]$ starting in the first panel (top left) with $\tilde\mu/2\tilde{t}=-1$ and ending in the last panel (bottom right) with $\tilde{\mu}/2\tilde{t}=1$.  For (a) and (b), the asymmetric character of the MZM amplitudes across the variation of $\tilde{\mu}$ correlates with the behavior of $\Delta_M$. It is suppressed for the negative $\tilde{\mu}$ region close to $\tilde{\mu}/2\tilde{t}=-1$. (a) has the same Dirac points as the Kitaev chain (c), the windows of maximum amplitude MZM are suppressed due to the long range interactions with and without imbalance. In (b) the maximum amplitude is strongly suppressed but finite.  Parameters are:  $N_s=40$ and $|\tilde{\Delta}_l|=2\tilde{t}$, $\tilde\Delta_p\approx\tilde\Delta_l$.
    }
    \label{fig:zoom_edge}
\end{figure*}

 First, I consider the cavity induced long-range interaction with a short range imbalance given by $g_p\neq0$. Indeed, it is found that the MZM are localized at the edges, as  the mass indicator $\Delta_M$ vanishes when the number of sites increases, away from $|\tilde{\mu}/2\tilde{t}|=1$, see Fig. \ref{fig:massind}. This has to do with the fact that the system only has three Dirac points, the same as the Kitaev chain.  The mass indicator for the short range Kitaev chain can be computed analytically as $\Delta_M/|\tilde{\Delta}_p|=\tilde{\mu}/2\tilde{t}(-\tilde{\mu}/2\tilde{t})^{N_s-1}$ when $|\tilde{\Delta}_p|=2\tilde{t}$ ~\cite{Pupillo2,Fendley}, symmetric in the speed of convergence to reach $\Delta_M$=0 around $\tilde{\mu}=0$ as the number of sites increases. 
After performing numerical simulations for $\mathcal{H}_{\mathrm{eff}}$ with $g_p\neq0$~\cite{fn2}, the results can be fitted to,
 \begin{equation}
 \frac{\Delta_M}{|\tilde{\Delta}_l|}\approx-\frac{9}{10N_s}+\sum_{n=1}^9\frac{(-1)^{n+1} \alpha_n}{N_s^{1+\delta_n}}\left(\frac{\tilde{\mu}}{2\tilde{t}}\right)^n,
 \label{dm}
 \end{equation}
 with
  $\alpha_n\approx (6.16-11.93 n+ 8.9n^2-0.6n^3)/n^{0.82}$ 
  and 
  $\delta_n\approx -(17+2n^2)\times10^{-3} $ 
  for $-0.6\lesssim\tilde{\mu}/2\tilde{t}\lesssim0.6$ and $\tilde{\Delta}_p/2\tilde{t}\sim1$, other choices present a similar series structure. 
 Thus, even though the decay is not as fast  as for the Kitaev chain, the MZM will appear for $V_p\neq 0$ and $g_p\neq0$ for large number of sites as the systems moves away from $|\tilde{\mu}/2\tilde{t}|=1$.  $\Delta_M$ decreases faster for $0<\tilde{\mu}/2\tilde{t}<1$ compared to $-1<\tilde{\mu}/2\tilde{t}<0$. In Fig. \ref{fig:massind}, I show the maximum difference between mass indicators at $\tilde{\mu}/2\tilde{t}=-0.6$ (slower) and $\tilde{\mu}/2\tilde{t}=0.6$ (faster) as the number of sites increases. The imbalance effect can be traced back to the fact that the combination with the long range interaction  produces different signs in (\ref{dm}) as the system moves away from $\tilde\mu=0$. This effect originates as additional sites can contribute to the low energy mode dynamics at the level of the effective equations of motion~\cite{Fendley}. Therefore, it will be easier to observe MZM for $0\leq\tilde{\mu}<2\tilde{t}$ when short range imbalance is introduced to $\mathcal{H}_{\mathrm{eff}}$. 
The situation changes dramatically when $\mathcal{H}_{\mathrm{eff}}$ is considered with $g_p=0$,. Following the same technique,  instead of convergence towards finding edge modes the signal for the edge modes ($|\Delta_M|=0$), $|\Delta_M|$ oscillates vigorously for $|\tilde{\mu}|<2\tilde{t}$, with the frequency increasing as $N_s$ increases. I interpret this as a signature of the fact that now the edge modes are not maximal on the whole range of $\tilde\mu$. So, while there exists points for $|\Delta_M|=0$, these are not indicative of stable windows for the appearance of edge modes, see Fig. \ref{fig:dpsi_cavs}. The mass gap indicator gets scrambled because additional interference is manifest as the gap between the zero modes and the bulk is much smaller, as we will see below.

{\it Majorana windows}. In order to confirm the above,  I employ the methods shown in~\cite{Frank} to extract the amplitude of the edge modes via singular value decomposition (SVD) of  $\mathcal{H}_{\mathrm{eff}}$. Using SVD is it is found that the amplitude of the edge modes is partially suppressed between the modes at the two extrema of the system when $g_p=0$. I interpret this imbalance as the signal for the emergence of accumulation of amplitude at the edges of the system, i.e. MZM. However, in contrast to the case of  $g_p\neq0$ these are not maximum amplitude MZM in the typical sense, i.e. the amplitudes do not maximize at the edges reaching unity, but there is a a measurable imbalance with partial suppression of all other amplitudes (sites different from the edges). Looking closely at the formation of the edge modes, I compute the amplitude of the edge modes for the corresponding regions in between Dirac points given by $q^D_{\mathrm{cav}}(n)$, the chemical potential corresponding to the emergence of a possible MZM at each Dirac point is  $\tilde{\mu}^D_n=-2\tilde{t}\cos(q^D_{\mathrm{cav}}(n))$. Therefore, at each interval $[\tilde{\mu}^D_n,\tilde{\mu}^D_{n+1}]$ for $n\leq N_s+1$ there is the possibility to have MZM. However, the multiplicity of the Dirac points across the whole interval $\tilde\mu<2 t$ complicates the analysis with respect to the Kitaev model.  In the Kitaev model one can use the behavior of the Pfaffian between the Dirac points to compute the edge mode or Majorana invariant~\cite{Kitaev}, for $V_p\neq 0$ and $g_p=0$ this does not work. This is because it can predict absence or presence depending if the number of Dirac points is even or odd and it doesn't consider the damped nature of the amplitude, rendering it ambiguos. To verify the emergence of isolated edge modes, I compute directly the distance $D\psi=||\Delta\psi_{R/L}|-\bar\psi_{R/L}|$, which is the difference of amplitude of the edge modes and the mean value of the amplitude over all sites. In the panels of Fig. \ref{fig:zoom_edge},  I show $D\psi$ between all the intervals of the chemical potential starting for $|\tilde{\mu}|\leq 2\tilde{t}$.  Moving away from $|\tilde{\mu}|=2\tilde{t}$, there are windows with standard edge modes (amplitude is maximal) with several regions of ``oscillating suppressed amplitude edge modes"  in between. Interestingly, the regular way to define the Majorana invariant  does not yield conclusive evidence to identify the aforementioned windows. In the simulations it is found that sometimes counterintuitively the stability of the edge mode can extend  between Dirac points  (maximal between more than one panel in Fig. \ref{fig:zoom_edge}) or be partially suppressed in a region of the interval (maximal in just part of the interval in Fig. \ref{fig:zoom_edge}).

\begin{figure}
    \centering
    \includegraphics[width=0.45\textwidth]{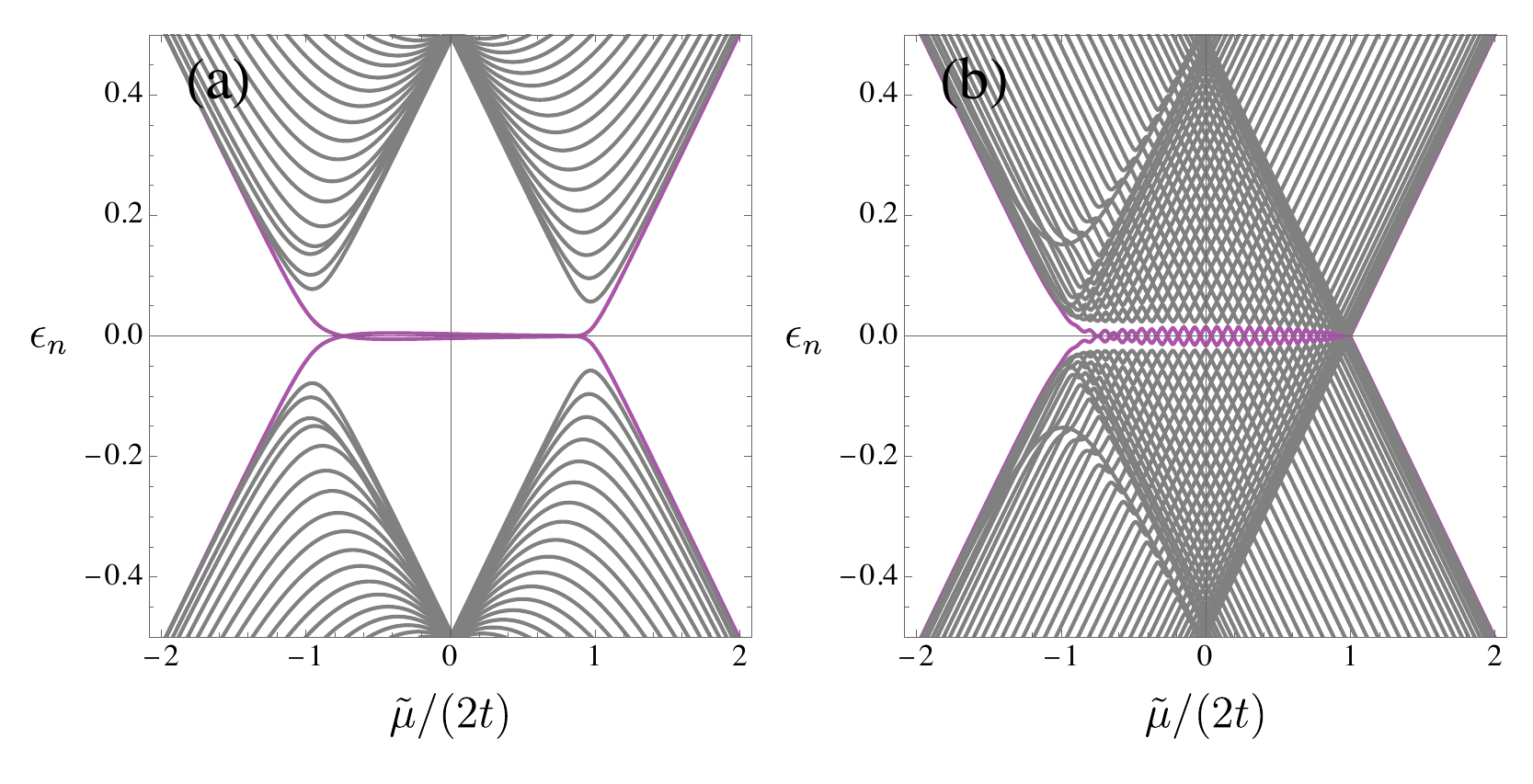}
    \caption{{\bf Spectrum.} The bulk is in gray with the lowest magnitude modes in purple. (a): $V_p\neq0$, $g_p\neq0$ , where it  can be identified there are to a very good approximation MZM in the range $|\tilde{\mu}|<2|\tilde{t}|$. (b): $V_p\neq0$, $g_p=0$, there are oscillations in the amplitude of the modes, thus these are not maximum amplitude edge modes but there are values where they can have zero energy while the gap is strongly suppressed in the region $|\tilde{\mu}|<2|\tilde{t}|$. Parameters are:  Energies are in units of $\tilde t$ for $N_s=40$ and  $|\tilde{\Delta}|=2\tilde{t}$.
    }
    \label{fig:sp_cavs}
\end{figure}

{\it Spectrum and MZM}. The interplay between short range and long range interactions stabilizes the edge modes, consistent with the mass indicator, see Fig. \ref{fig:dpsi_cavs} (a) for $g_p\neq 0$ and $V_p\neq0$. This correlates with the behavior of the lowest magnitude modes in the spectrum Fig. \ref{fig:sp_cavs} (a). In contrast, the energy gap to the bulk mode of the lowest magnitude modes in the case of  $V_p\neq0$ and $g_p=0$ is suppressed  Fig. \ref{fig:sp_cavs} (b), the behaviour of the MZM in the same intervals can be seen in  Fig. \ref{fig:zoom_edge}. However, there are points in the $|\tilde{\mu}|\leq 2\tilde{t}$ interval that one can have zero energy modes Fig. \ref{fig:zoom_edge} and Fig. \ref{fig:sp_cavs} (b). The system with cavity induced interactions $\mathcal{H}_{\mathrm{eff}}$  have always suppressed windows of large amplitude MZM with respect to the Kitaev chain $V_p=0$ and $g_p\neq0$ for the same number of sites and in the same interval windows, Fig. \ref{fig:zoom_edge}. With the results from the spectrum, the mass gap and the direct computation of edge mode amplitudes it can be confirmed that indeed short range interactions can maximize the amplitude of MZM. In an experiment, the emergence of MZM could be measured directly by in-situ imaging the accumulation of density at the edges of the chain or by measuring the light in the cavity, as the number of photons is $n_{\mathrm{ph}}=\langle \hat a^\dagger \hat a\rangle\approx|\tilde{g}_{\mathrm{eff}}/\Delta_c|(|\Delta_l|^2+\langle\hat n\rangle^2/2)N_s$, when the light and matter couple, typical of the superradiant response~\cite{P}, see Fig. \ref{fig:system}. The collective light emission and the presence of the edge modes would confirm a topological superradiant fermionic phase of matter when measured.

Summarizing, I have found that MZM are present with cavity induced interactions and these can be stabilized (maximized in amplitude) with the help of short range interactions leading to similar behaviour as the MZM in the Kitaev chain. Thus, the interplay with short range interactions offers a degree of control that can help to open the energy gap to the bulk spectrum to maximize the amplitude of the MZM. Therefore, the superradiant states that emerge in the system are topological as they support MZM. In general, we expect that these MZM could be probed to be used perhaps for quantum information purposes~\cite{QIA1,QIA2,QIA3}, as using the geometric arrangement of the system (cavity and optical lattice axes) can provide sections where the pairing could be induced or suppressed as the light is pumped into the system~\cite{PRAW}, while the cavity-pump detunning provides yet another control parameter with some freedom. With the help of the light coupling one could draw the wires with pairing with the light and assemble the structures needed to have useful q-bits analogous to \cite{QIA1} in a single setup while modifying the structures on demand. Moreover, the methods that were developed to find the effective ground state of the model can be used to study similar fermionic and analogous systems incorporating the Majorana representation of the Hamiltonian and Light-Matter self-consistency. 

The phenomenology described in this letter could be explored readily in  $^6$Li system ~\cite{CF1,CF2,CF3, CF4} or in a yet to be implemented $^{40}$K system inside a high-Q cavity with the addition of an optical lattice  and spin polarized fermions with current experimental setups. Ultracold atoms with an optical lattice inside a high-Q cavity  have already been achieved in bosonic systems~\cite{E1,E2,E3,H1,P}.  Perhaps, similar behaviour could be expected in analog light-induced superconducting systems ~\cite{FL1,FL2,FL3,FL4,FL5} with elongated geometry arrays in some limit.

\begin{acknowledgements}
I thank F. Mivehvar, T. Donner, J. Hofmann, H. Ritsch, P. Christodoulou and R. Guti\'errez-J\'auregui for useful discussions. This work is partially supported by the grants UNAM-DGAPA-PAPIIT: IN118823, UNAM-DGPA-PAPIIT: IG101826, UNAM-CIC: Apoyo a Laboratorios Nacionales 2025, CONAHCYT/SECIHITI: LNC-2023-51,  as well as by grants from NVIDIA and utilized NVIDIA RTX 6000 Ada.
\end{acknowledgements}


\begin{thebibliography}{99}
\bibitem{limit}For large number of sites the maximum contributions of $f(q)^2$ are: 
$\max[f(q_{\mathrm{max}})^2)]\approx \frac{2}{3} \sin ^4\left(\sqrt{\frac{3}{2}}\right)\approx 0.52$ and $q_{\mathrm{max}}\approx\pm \frac{2 \sqrt{3}}{\sqrt{2 N_s^2-1}}$.
\end{thebibliography}
\end{document}